# An Experimental and Theoretical Investigation of the Gas-Phase C($^3$P) + N$_2$O Reaction. Low Temperature Rate Constants and Astrochemical Implications

Kevin M. Hickson,[1,*] Jean-Christophe Loison,[1] Pascal Larregaray,[1] Laurent Bonnet[1] and Valentine Wakelam[2]

[1] Univ. Bordeaux, CNRS, Bordeaux INP, ISM, UMR 5255, F-33400 Talence, France

[2] Univ. Bordeaux, CNRS, LAB, UMR 5804, F-33270 Floirac, France

**Abstract**

The reaction between atomic carbon in its ground electronic state, C($^3$P), and nitrous oxide, N$_2$O, has been studied below room temperature due to its potential importance for astrochemistry, with both species considered to be present at high abundance levels in a range of interstellar environments. On the experimental side, we measured rate constants for this reaction over the 50-296 K range using a continuous supersonic flow reactor. C($^3$P) atoms were generated by the pulsed photolysis of carbon tetrabromide at 266 nm and were detected by pulsed laser induced fluorescence at 115.8 nm. Additional measurements allowing the major product channels to be elucidated were also performed. On the theoretical side, statistical rate theory was used to calculate low temperature rate constants. These calculations employed the results of new electronic structure calculations of the $^3$A″ potential energy surface of CNNO and provided a basis to extrapolate the measured rate constants to lower temperatures and pressures. The rate constant was found to increase monotonically as the temperature falls ($k_{C(^3P)+N_2O}$(296 K) = (3.4 ± 0.3) × 10$^{-11}$ cm$^3$ s$^{-1}$), reaching a value of $k_{C(^3P)+N_2O}$(50 K) = (7.9 ± 0.8) × 10$^{-11}$ cm$^3$ s$^{-1}$ at 50 K. As current astrochemical models do not include the C + N$_2$O reaction, we tested the influence of this process on interstellar N$_2$O and other related species using a gas-grain model of dense interstellar clouds. These simulations predict that N$_2$O abundances decrease significantly at intermediate times (10$^3$ – 10$^5$ years) when gas-phase C($^3$P) abundances are high.

## 1 Introduction

As the fifth most abundant element in the Universe, nitrogen and its compounds are ubiquitous throughout the interstellar medium (ISM). In particular, nitrogen-bearing organic species such as HCN and $CH_3CN$ are amongst the most commonly detected of all molecules, across a diverse range of objects and over a wide range of temperatures. In contrast, only four neutral nitrogen oxide species (two stable molecules NO, $N_2O$ and two unstable species HNO and HONO) have been positively identified in the interstellar medium (ISM).[1] While NO was detected in 1978,[2] $N_2O$ was first detected much more recently through its J = 3 → 2, 4 → 3, 5 → 4 and 6 → 5 rotational transitions by Ziurys et al.[3] towards Sagittarius B2 (M) (Sgr B2 (M)), with a fractional abundance of $10^{-9}$ (all quoted observational abundances are relative to $H_2$). Subsequent $N_2O$ observations by Halfen et al.[4] centered on the Sgr B2 (N) core derived an $N_2O$ relative abundance of ~ $1.5 \times 10^{-9}$. More recently, Ligterink et al.[5] detected $N_2O$ through its J = 14 → 13 transition towards the low-mass solar-type protostellar binary IRAS 16293–2422B with a derived relative abundance of $3.3 \times 10^{-9}$, while Agundez et al.[6] observed this molecule through its J = 4 → 3 transition in L483, a dense core surrounding a class 0 protostar, deriving a relative abundance of $1.45 \times 10^{-10}$. As the observed $N_2O$ abundance levels are quite high, it is important to consider its role in the overall chemistry of the ISM. In terms of potential coreagent species, neutral carbon, nitrogen and oxygen atoms are all considered to be present at high abundance levels in the dense ISM ($10^{-4}$ – $10^{-5}$), making it particularly important to evaluate their reactivity towards other interstellar species. While the neutral-neutral reactions of C- and N-atoms with NO molecules are reasonably well characterized at low temperature through several experimental and theoretical studies[7-15] (the reaction between ground state atomic oxygen and nitric oxide occurs through three body association[16] which will be negligible at the densities of interstellar clouds), there are very few studies of gas-phase $N_2O$ reactions at interstellar temperatures. In terms of its reactivity towards atomic radicals, $N_2O$ is not expected to react with N- or O-atoms at low temperature (the $N + N_2O$ reaction is spin forbidden[17] while the $O + N_2O$ reaction is characterized by a large activation barrier[18]). In contrast, the reactivity of atomic carbon towards $N_2O$ has been studied at room temperature by the Husain group[19, 20] using the flash photolysis of carbon suboxide as a source of ground state atomic carbon with these atoms being detected by resonance absorption around 166 nm. Interestingly, they derived large rate constants of $(2.6 \pm 1.3) \times 10^{-11}$ cm$^3$ s$^{-1}$ and $(1.3 \pm 0.3) \times 10^{-11}$ cm$^3$ s$^{-1}$ respectively, while the later measurements of Dorthe et al.[21] using the microwave discharge of CO as a source of carbon atoms coupled with the chemiluminescence detection of electronically excited NO products yielded a slightly smaller room temperature value for the rate constant of $(8.5 \pm 1.6) \times 10^{-12}$ cm$^3$ s$^{-1}$. The relatively large values measured for this process at 300 K indicate that it might occur in the absence of an activation barrier. Nevertheless, there are no direct theoretical studies of this system in the literature to validate this hypothesis or experimental studies covering a range of temperatures. At the present time, the $C + N_2O$ reaction is absent from all astrochemical databases so its influence on interstellar chemistry is currently unknown.

To address this issue, and to evaluate the potential importance of the reaction between C-atoms and $N_2O$ molecules in interstellar space, we performed a joint experimental and theoretical study. On the experimental side, a supersonic flow reactor was employed coupled with pulsed laser photolysis to produce atomic carbon while laser induced fluorescence was used to follow the kinetics of C-atom loss in the presence of $N_2O$ between 50 and 296 K. On the theoretical side, electronic structure calculations were performed at various levels (MRCI+Q, DFT and CCSD(T)) to obtain a clearer picture of the overall mechanism of the C + $N_2O$ reaction. In particular, we investigated the various reaction pathways and determined the relevant intermediates, transition states and complexes over the potential energy surfaces of NCNO adiabatically connecting reagents to products. Additionally, Rice-Ramsperger-Kassel-Marcus / master equation (RRKM/ME) calculations were performed[22] to calculate the rate constants for this reaction and extrapolate the experimental data to lower temperatures and pressures.

Finally, we tested the potential effects of the C + $N_2O$ reaction on interstellar chemistry by including this process in a state of the art gas-grain astrochemical model. Section 2 describes the experimental techniques employed, section 3 outlines the methodology used in the electronic structure calculations and section 4 details the rate constant calculations. Section 5 presents the experimental and theoretical results and discusses these results in the context of earlier work. The astrophysical implications of the current work and our conclusions are presented in sections 6 and 7 respectively.

**2 Experimental Methods**

The experimental work reported here was performed using the CRESU technique (Cinétique de Reaction en Ecoulement Supersonique Uniforme or Reaction Kinetics in a Uniform Supersonic Flow), employing an existing CRESU apparatus which has been previously described.[23, 24] Numerous modifications have been applied to the original system to allow the kinetics of reactions involving atoms in their ground ($N(^4S)$,[12, 25-28] $C(^3P)$[29-31]) and excited ($C(^1D)$,[32-39] $N(^2D)$,[40-42] $O(^1D)$[37, 43-48]) electronic states to be investigated at temperatures relevant to planetary atmospheres and the interstellar medium. Experiments were performed at five different temperatures (50 K, 75 K, 127 K, 177 K and 296 K), making use of several different Laval nozzles to generate low temperature flows during the course of this study. The flow characteristics of these nozzles are listed in Table 2 of Hickson et al.[30] (the nozzle generating flows at 106 K was not used in the present work). Room temperature experiments were conducted by using the reactor as a conventional flash-photolysis apparatus under slow flow conditions.

C($^3$P) atoms were produced in the supersonic flow by the multiphoton dissociation of carbon tetrabromide (CBr$_4$). To introduce these precursor molecules into the supersonic flow, a small flow of the carrier gas was passed over solid CBr$_4$ contained in a separate vessel maintained at a known pressure and room temperature. The maximum CBr$_4$ concentration used in the cold supersonic flow during these measurements was estimated to be $2.6 \times 10^{13}$ cm$^{-3}$ based on its saturated vapour pressure. The fourth harmonic output of a 10 Hz pulsed Nd:YAG laser at 266 nm (22 mJ/pulse) was coaligned along the supersonic flow so that the nascent C($^3$P) concentration produced was essentially identical at any axial position due to the negligible attenuation of the UV photolysis beam at this wavelength ($\sigma_{CBr_4}(266 \text{ nm}) = 1 \times 10^{-18}$ cm$^2$).

In earlier experiments,[29-31] C($^3$P) atoms were detected directly by pulsed resonant laser induced fluorescence in the vacuum ultraviolet (VUV LIF) using the 2s$^2$2p$^2$ $^3$P$_2$ → 2s$^2$ 2p3d $^3$D$_3$° transition at 127.755 nm. Here a different transition was employed, namely the 2s$^2$2p$^2$ $^3$P$_2$ → 2s$^2$ 2p5d $^3$D$_3$° one at 115.803 nm. Although the emission transition probability was slightly weaker for this line, it provided the possibility to follow several atomic species (notably C($^3$P), O($^1$D) and N($^2$D)) without having to change the laser dye and the associated beam-steering optics. Tuneable narrowband radiation at 115.8 nm was generated from a 10 Hz Nd:YAG pumped dye laser operating around 695 nm. The fundamental dye laser beam was frequency doubled in a beta barium borate (BBO) crystal, with the residual dye laser radiation being separated from the 347 nm UV beam by two dichroic mirrors with a coating optimized for reflection at 355 nm. The 347 nm beam was then directed and focused into a cell attached at the level of the observation axis containing 50 Torr of xenon with 160 Torr of argon added for phase matching purposes. The third harmonic VUV beam generated by this procedure was collimated at the cell exit by a $MgF_2$ lens before entering a 75 cm long sidearm separating the tripling cell from the reactor. This sidearm contained a series of diaphragms to block most of the divergent UV beam before it could enter the reactor and was constantly flushed with nitrogen or argon to prevent additional attenuation of the VUV beam by reactive gases. On entering the reactor, the VUV beam was positioned to intersect the cold supersonic flow at right angles, with the detector similarly positioned at right angles to both the flow and the VUV beam to minimize the detection of scattered VUV and UV radiation. The detection system itself consisted of an LiF lens which focused the fluorescence emission from unreacted carbon atoms within the supersonic flow onto the photocathode of a solar blind photomultiplier tube (PMT). The PMT was isolated from the reactor by a LiF window, while the region between the PMT and the reactor was evacuated to prevent attenuation of the VUV emission by atmospheric oxygen. As with the VUV excitation source, several diaphragms were placed throughout the detection system to minimize the detection of scattered light. VUV LIF signals from $C(^3P)$ atoms were recorded as a function of time between the photolysis and probe lasers using a boxcar integration system, with the delay being controlled by a digital delay generator. 30 laser shots were averaged at each time point with decay curves consisting of at least 70 time points that were generally measured until at least six half-lives had elapsed (with the exception of decays recorded in the absence of coreagent molecules). For each series of measurements, the Laval nozzle was maintained at a fixed distance from the observation axis, corresponding to the maximum displacement for which flow conditions could be considered optimal. These values were derived in earlier characterization experiments performed using a Pitot tube to record the impact pressure of the supersonic flow as a function of distance. The coreagent $N_2O$ (Sigma Aldrich 99 %) and carrier gas Ar (Linde 99.999 %) and $N_2$ (Air Liquide 99.999 %) flows were regulated by calibrated mass-flow controllers. All gases including the Xe (Linde 99.999 %) used in the tripling cell were flowed directly from gas cylinders without further purification.

**3 Electronic Structure Calculations**

The potential energy surface (PES) of the C + $N_2O$ reaction has been studied at different levels of theory. Firstly, the global pathways connecting the reagents C and $N_2O$ with the various possible products were calculated using DFT calculations with the M06-2X functional[49] associated to the aug-cc-pVTZ (AVTZ) basis set. Secondly, more accurate calculations were performed on the entrance valley of the reaction using the RCCSD(T)-F12/AVTZ method. This method was also used to calculate the characteristics (geometry, energy and vibrations) of the most important stationary points. The entrance valley was also calculated using the Davidson corrected multi-reference configuration interaction (MRCI + Q) with complete active space self-consistent field (CASSCF) wave-functions associated to the AVTZ basis. The CASSCF and MRCI calculations were performed only to check the RCCSD(T)-F12 results and then using only a relatively small active space with 6 electrons distributed in 6 orbitals. The CCSD(T) and MRCI+Q calculations were performed using the MOLPRO 2010 package. DFT calculations were carried out with the Gaussian09 package.[50] The CASCCF calculations lead to mono-configurational wave-functions for adducts and transition states. CCSD(T) calculations should therefore be accurate as also indicated by the T1 and D1 diagnostic values (T1 values are in the 0.018 - 0.020 range and D1 values are in the 0.046 - 0.050 range). The schematic energy diagram for the triplet surface is shown in Figure 1. The calculated stationary point energies, geometries and frequencies of the C…N$_2$O complex and TSs are available in the supplementary information (SI) file.

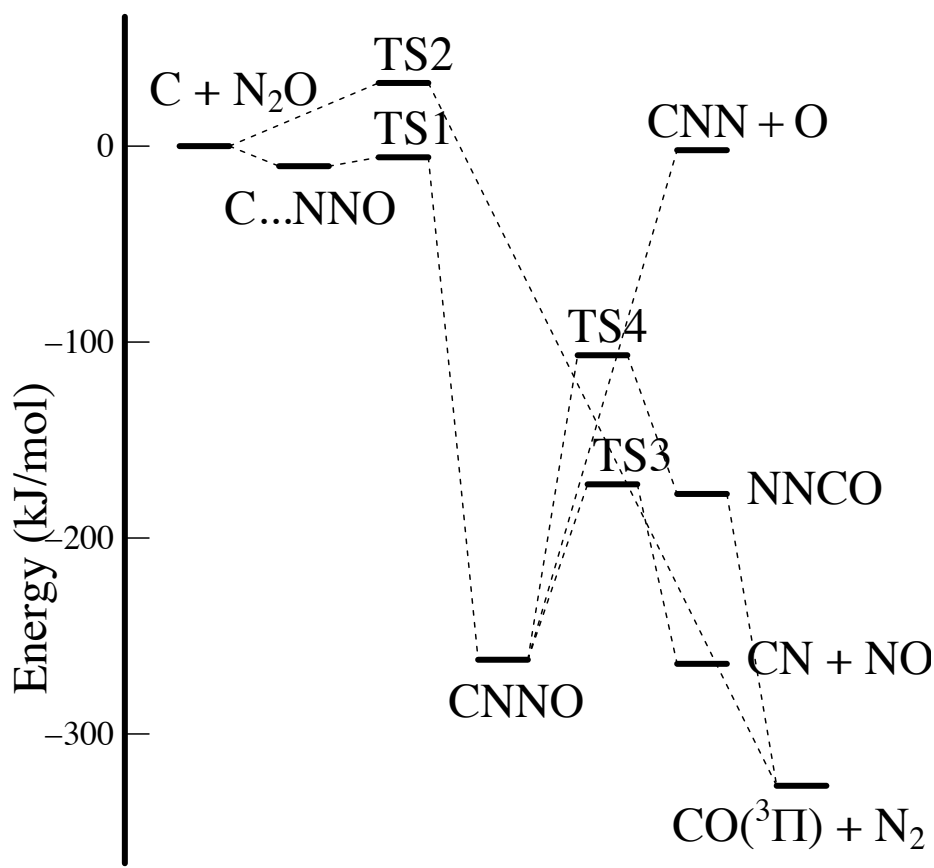

**Figure 1** Potential energy diagram for the C($^3$P) + N$_2$O reaction over the triplet state potential energy surface of NCNO calculated with the M06-2X functional at the aug-cc-pVTZ(AVTZ) level (the C…NNO complex, TS1 and TS2 energies represented here are those calculated at the CCSD(T)-F12/AVTZ level).

The C($^3$P$_{0,1,2}$) + N$_2$O($^1\Sigma^+$) system correlates with 3 surfaces, one $^3$A' and two $^3$A'' surfaces in C$_s$ symmetry. At the MRCI+Q/AVTZ level with geometry fixed at the C($^3$P$_{0,1,2}$) and N$_2$O($^1\Sigma^+$) geometry, and RCCSD(T)-F12-AVTZ levels with optimized geometry, only the first $^3$A'' surface is attractive for the approach of atomic carbon to the terminal nitrogen atom of N$_2$O, leading to the C…N$_2$O van der Waals complex, while the approach to the oxygen atom presents a barrier (TS2 in Figure 1). The calculated potential energy curve of the first $^3$A'' surface when carbon atom approaches N$_2$O is presented in the SI file. The C…N$_2$O complex, with the carbon atom attached to the terminal nitrogen atom, is weakly bound. Further evolution of the C…N$_2$O complex leads to CNNO in a triplet state through a submerged transition state (TS), denoted as TS1, 6.7 kJ/mol below the C + N$_2$O entrance level when ZPE effects are considered. TS1 is very close in energy to the C…N$_2$O complex at the M06-2X/AVTZ level (with an even smaller energy when ZPE effects are included) although this energy is notably higher, 3.01 kJ/mol below the C + N$_2$O entrance level when ZPE effects are considered at the RCCSD(T)-F12/AVTZ level. Considering the accuracy of the CCSD method and the good T1 and D1 parameters (0.021 and 0.050 respectively for the TS1 calculations), the RCCSD(T)-F12/AVTZ results are certainly more realistic. Then, the adduct CNNO can either give CN + NO via TS3 or isomerize to NNCO via a cyclic transition state TS4. NNCO is not a stable compound, although there seems to be a local minimum on the triplet surface, and this species should easily evolve to CO(a³Π) + N₂ (we could not find a transition state for this transformation, although this does not mean that it does not exist), nor did we find any transition states connecting CNNO and NNCO to NCNO and NCON on the triplet surface. Considering the energies of TS3 and TS4, and if reaction occurs predominantly on the triplet surface, CN + NO are likely to be the main reaction products with CO + N₂ products as a potentially non-negligible reaction pathway if the reaction behaves statistically. The results of supplementary experiments performed to quantify the respective product channels are described in section 5.2.

**4 Rate constant calculations**

As shown in Figure 1, the C + N₂O reaction can be well described by a two TS model where an outer TS is associated with the barrierless formation of the C…N₂O complex, while an inner submerged TS (TS1) is associated with the isomerization step from C…N₂O towards the CNNO adduct intermediate. All rate constant calculations were performed using the *Multiwell* program suite.[22] As a first step, rate constants for the unimolecular dissociation of the C…N₂O complex towards the C + N₂O reagents were calculated using microcanonical transition state theory (μCTST) as implemented in the *ktools* program of *Multiwell*.[51] The energy ($E$) and angular momentum ($J$) resolved microcanonical reaction rate constant, $k(E,J)$, is given by

$$k(E,J) = \frac{1}{h} \frac{G^{\neq}(E - E_{0,J}, J)}{\rho(E,J)}$$

where $h$ is Planck's constant, $G^{\neq}(E - E_{0,J}, J)$ is the sum of states of the TS, $E_{0,J}$ is the reaction critical energy and $\rho(E,J)$ is the density of states of the reagent molecule for a given $(E,J)$. As there is no intrinsic barrier for C…N₂O (re)dissociation, *ktools* applies a variational treatment to determine the rate constant.[52] For this purpose, a series of constrained optimizations were performed at fixed bond distances along the minimum energy path along the entrance channel between C…N$_2$O and the separated reagents at the RCCSD(T)-F12/AVTZ level. At each distance, the vibrational frequencies orthogonal to the reaction path, the rotational constants and the derived enthalpies of formation at 0 K were derived and used as inputs in the calculations. A small grain size of 1 cm$^{-1}$ was chosen to improve accuracy at low temperature. At each fixed distance, a trial rate constant was calculated, with the position of the variational transition state being identified by the point at which the minimum trial rate constant occurred. If more than one bottleneck was found to exist along the reaction path, the unified statistical theory developed by Miller[53] was applied to calculate the overall rate constant.

The values of $k(E,J)$ were then averaged over $E$ and $J$ at a given temperature[54] to yield the canonical rate constant, equivalent to the high-pressure-limit rate constant at a given temperature, $k_{\infty,\mu}^{uni}(T)$. The capture rate constant, $k_{capt}(T)$, then followed by application of the equilibrium constant, $K_{eq}(T)$, through the expression

$$k_{capt}(T) = k_{\infty,\mu}^{uni}(T)/K_{eq}(T)$$

The values of $k_{capt}(T)$ were then used as inputs for ME simulations to obtain temperature and pressure-dependent rate constants.

RRKM statistical theory as implemented in *Multiwell* was employed to derive energy-dependent microcanonical rate constants k($E$) for a unimolecular reaction

$$k(E) = L^{\neq} \frac{g_e^{\neq}}{g_e} \frac{1}{h} \frac{G^{\neq}(E-E_0)}{\rho(E)}$$

$L^{\neq} = \frac{m^{\neq}}{m} \times \frac{\sigma_{ext}}{\sigma_{ext}^{\neq}}$ is the reaction path degeneracy, where $m^{\neq}$ and $m$ are the number of optical isomers, $\sigma_{ext}^{\neq}$ and $\sigma_{ext}$ are the external rotation symmetry numbers, for the TS and reagent respectively. $g_e^{\neq}$ and $g_e$ are the electronic state degeneracies of the transition state and reactant, respectively. *h* is Planck's constant, $G^{\neq}(E - E_0)$ is the sum of states of the TS, $E_0$ is the reaction critical energy and $\rho(E)$ is the density of states of the reagent molecule.

Here, the C + N$_2$O bimolecular association reaction is assumed to result in the production of a chemically activated species C…N$_2$O* which can further evolve by: (I) redissociating to reactants (II) reacting further via TS1 (TS2 will only play a role at very high temperatures) (III) stabilizing to C…N$_2$O through collisions with the carrier gas. Only the pathway C + N$_2$O ⇋ C…N$_2$O → TS1 → CNNO → TS3 → CN + NO was considered to contribute significantly to the overall reaction rate, as indicated by product yield experiments described later on. The simulations were performed by starting from the chemically activated species C…N$_2$O*, with the rate constant for loss of C($^3$P) + N$_2$O reagents (the experiments measure total loss of C($^3$P)), $k(T, [M])$, given by

$$k(T, [M]) = k_{capt}(T)\,(1 - f(T, [M]))$$

where $f(T, [M])$ is the fraction of reagent molecules remaining at the end of the simulation.

The simulations were performed over a wide range of pressures (100 kPa - $1 \times 10^{-8}$ Pa) and temperatures (300 K - 10 K). The use of very low pressures and temperatures allowed us to evaluate the reactive rate constants under conditions well beyond those accessible during the measurements, and relevant to interstellar environments. During the present simulations, in a similar manner to the $k_{capt}(T)$ calculations, a grain size of 1 cm$^{-1}$ was employed to ensure better precision at low temperatures. Lennard-Jones parameters were estimated for the C…N$_2$O and CNNO species ($\sigma$ = 4.5 Å, $\varepsilon$ / k$_B$ = 550 K), while values for the carrier gases Ar and N$_2$ were taken from the literature (Ar $\sigma$ = 3.3 Å, $\varepsilon$ / k$_B$ = 144 K; N$_2$ $\sigma$ = 3.7 Å, $\varepsilon$ / k$_B$ = 85 K).[55] Energy transfer was described using the standard exponential-down model.[56] Sums and densities of states for the C…N$_2$O vdW complex, CNNO, TS1 and TS3 were calculated using the *Densum* code which employs the Stein-Rabinovitch[57] version of the Beyer-Swinehart algorithm[58] for exact counts of states for species comprised of separable degrees of freedom. In contrast, as *Densum* is only suitable for fixed transition states, sums of states for the variational outer TS were derived by the *ktools* program instead, during calculations of $k_{capt}(T)$. Enthalpies of reaction ($\Delta H_{rxn}(0\ K)$) were supplied, in addition to the zero-point corrected reaction critical energies (barrier heights with ZPE correction). Additionally, although it was considered unlikely that tunneling would be important for this system, given the low value of the TS1 barrier, tunneling effects were nonetheless accounted for in the present simulations through the use of a one dimensional unsymmetrical Eckart barrier.[59]

## 5 Results and Discussion

### 5.1 Rate constant measurements and calculations

A large excess concentration of $N_2O$ was used for all the experiments performed here, allowing the kinetic data to be analysed using the pseudo-first-order approximation. Consequently, the $C(^3P)$ atom VUV LIF signal decayed exponentially as a function of time according to the formula $I = I_0 \exp(-k_{1st}t)$, where $I$ and $I_0$ are the time-dependent and initial C-atom fluorescence intensities respectively and $k_{1st}$ is the pseudo-first-order rate constant for C-atom loss. Representative decay curves recorded at 75 K are shown in Figure 2.

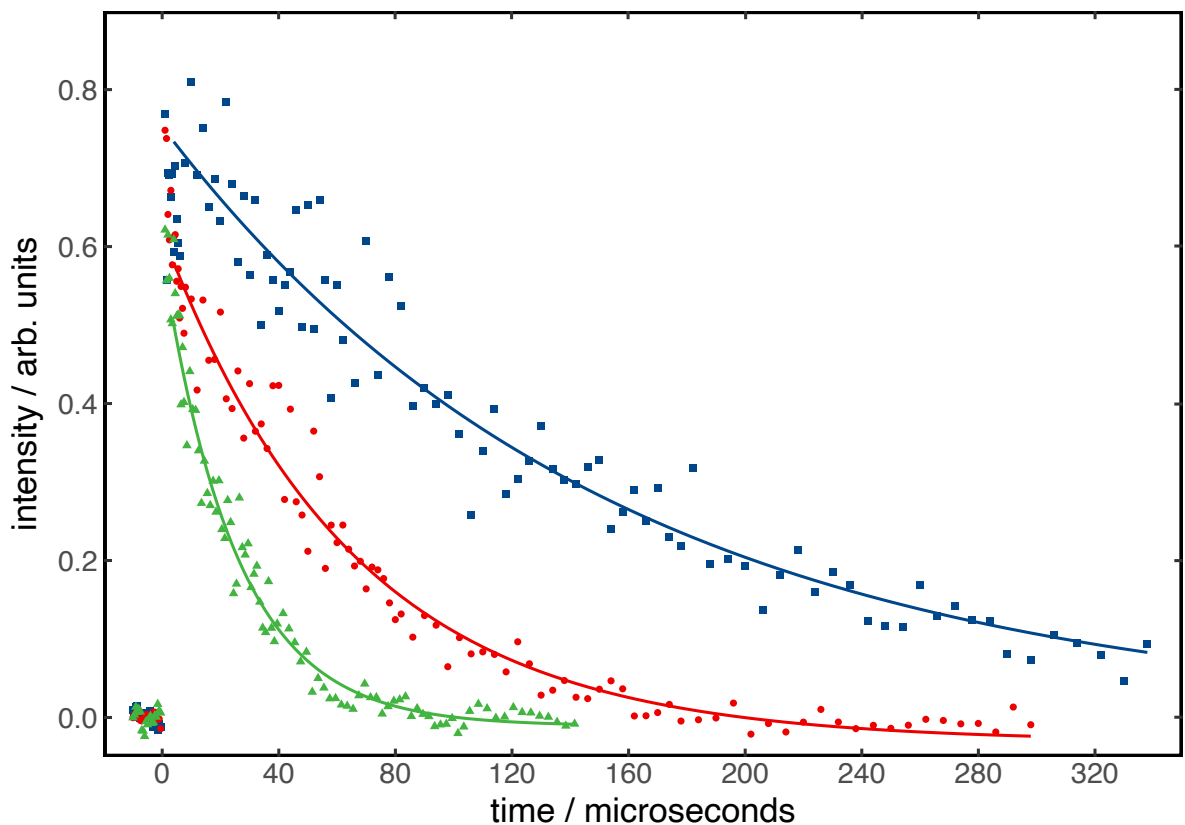

**Figure 2** C($^3$P) VUV LIF signal as a function of time recorded at 75 K. (Red solid circles) [N$_2$O] = 1.7 × 10$^{14}$ cm$^{-3}$ ; (green solid triangles) [N$_2$O] = 5.5 × 10$^{14}$ cm$^{-3}$ ; (blue solid squares) without N$_2$O. Solid lines represent exponential fits to the individual datasets.

In the absence of coreagent N$_2$O, C($^3$P) atoms are lost through diffusion from the observation zone (with a first-order rate constant $k_{diff}$) in addition to any secondary losses such as those arising from reaction with CBr$_4$ precursor molecules and any carrier gas impurities (given mostly by the term $k_{C+CBr_4}[CBr_4]$) so that $k_{1st} = k_{diff} + k_{C+CBr_4}[CBr_4]$ under these conditions. When N$_2$O is added to the supersonic flow, a large increase in the C($^3$P) loss rate is observed, a clear indication of the substantial reactivity between these two species below room temperature ($k_{1st} = k_{diff} + k_{C+CBr_4}[CBr_4] + k_{C+N_2O}[N_2O]$ under these conditions). For any particular temperature, kinetic decays were recorded for at least 6 different N$_2$O concentrations. Plots of the derived $k_{1st}$ values against the corresponding N$_2$O concentration, such as those displayed in Figure 3 for measurements at 50 K, 177 K and 296 K, yield the second-order rate constant from weighted linear least-squares fits to the data.

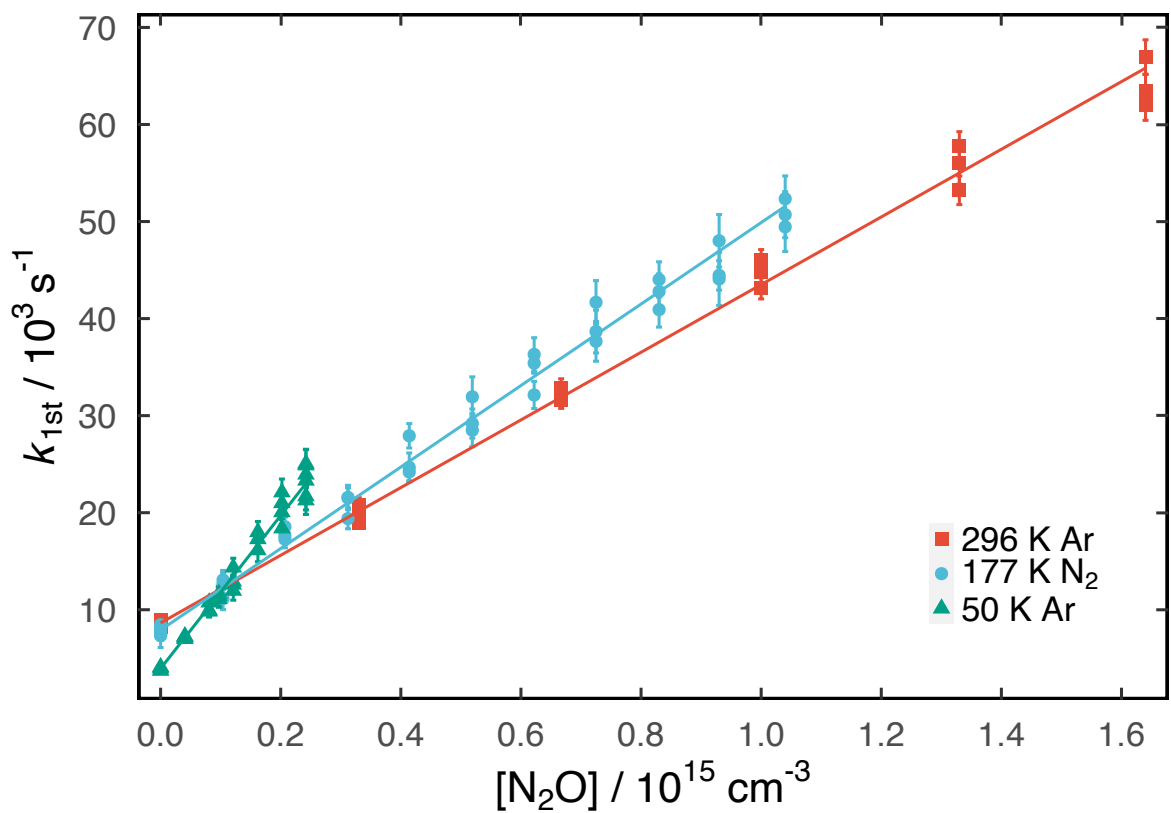

**Figure 3** Derived pseudo-first-order rate constants as a function of [N$_2$O]. (Green solid triangles) data recorded at 50 K in Ar; (blue solid circles) data recorded at 177 K in N$_2$; (red solid squares) data recorded at 296 K in Ar. Solid lines represent weighted linear least-squares fits to the individual datasets (weighted by the uncertainties associated to each datapoint, U, using the formula 1/U$^2$). The uncertainties themselves were derived from exponential fits such as those presented in Figure 2.

Here, the y-axis intercept values of the fits represent the sum of the C-atom secondary loss terms ($k_{diff}$ + $k_{C+CBr_4}[CBr_4]$) that do not vary with [N$_2$O], while the error bars on individual data points are derived from fits to temporal profiles such as those shown in Figure 2. The y-axis intercept value at 50 K is smaller than those derived at higher temperature due to two different effects. Firstly, [CBr$_4$] is lower due to a higher dilution factor at this temperature, so that the contribution from the $k_{C+CBr_4}[CBr_4]$ term is reduced. Secondly, the 50 K nozzle flow density is high (2.6 × 10$^{17}$ cm$^{-3}$) so that C-atoms diffuse more slowly, leading to a smaller contribution from $k_{diff}$. The measured second-order rate constants are listed in Table 1 with other relevant information and are displayed as a function of temperature in Figure 4 alongside earlier work and the present theoretical calculations.

**Table 1** Measured second-order rate constants for the C($^3$P) + N$_2$O reaction

| T / K | N[b] | [N$_2$O] / 10$^{14}$ cm$^{-3}$ | $k_{C(^3P)+N_2O}$ / 10$^{-11}$ cm$^3$ s$^{-1}$ | Carrier gas |
|---|---|---|---|---|
| 296 | 18 | 0 - 16.1 | (3.24 ± 0.32)[c] | N$_2$ |
| 296 | 18 | 0 - 16.4 | (3.49 ± 0.34) | Ar |
| 177 ± 2 | 36 | 0 - 10.4 | (4.20 ± 0.41) | N$_2$ |
| 127 ± 2 | 37 | 0 - 19.3 | (5.93 ± 0.60) | Ar |
| 75 ± 2 | 34 | 0 - 5.8 | (6.86 ± 0.66) | Ar |
| 50 ± 1 | 30 | 0 - 2.4 | (7.87 ± 0.80) | Ar |

[a]Uncertainties on the calculated temperatures represent the statistical (1σ) errors obtained from Pitot tube measurements of the impact pressure. [b]Number of individual measurements.

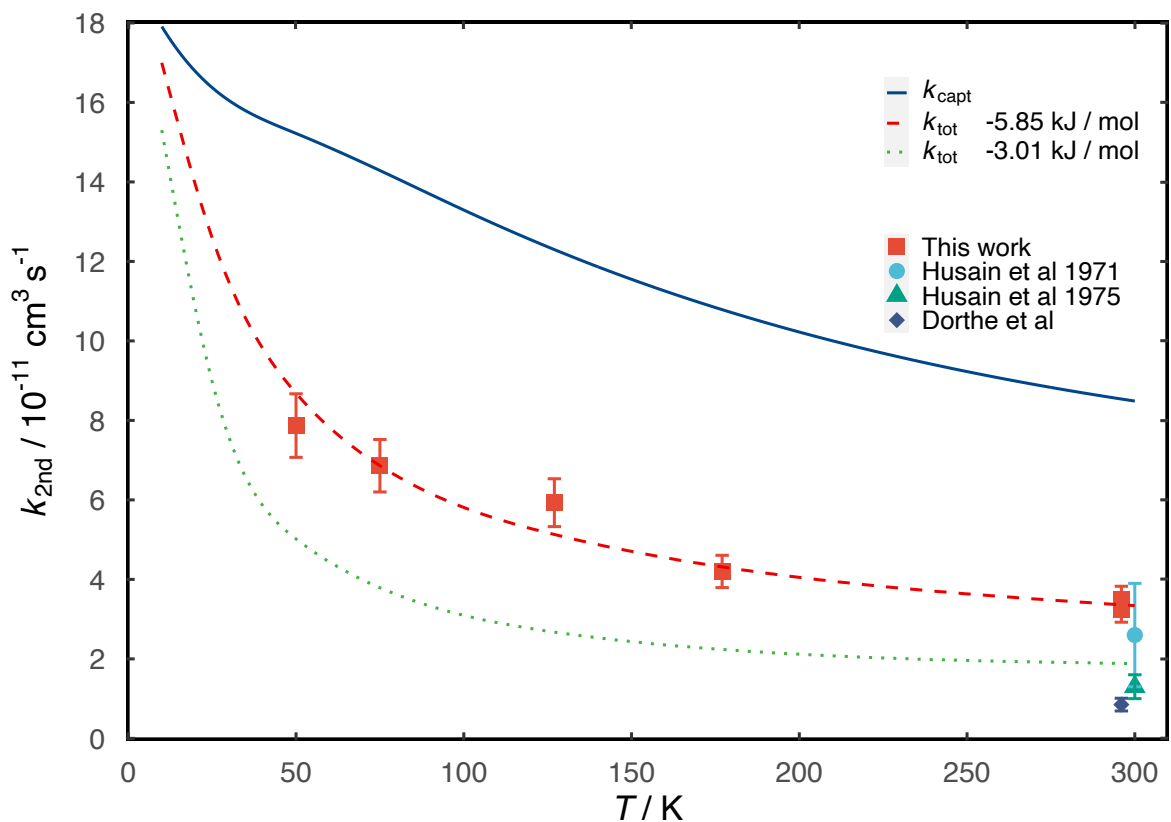

**Figure 4** Measured and calculated rate constants for the C($^3$P) + N$_2$O reaction as a function of temperature. Experimental studies: (blue solid diamonds) Dorthe et al.[21] ; (green solid triangles) Husain and Young[20] ; (blue solid circles) Husain and Kirsch[19] ; (red solid squares) this work. Theoretical studies: (solid blue line) high-pressure limit rate constant, $k_{capt}$ ; (dotted green line) overall rate constant considering the calculated submerged barrier of - 3.01 kJ/mol obtained at the RCCSD(T)-F12/AVTZ level with respect to the reagent asymptote; (dashed red line) overall rate constant after empirical adjustment of the calculated submerged barrier to -5.85 kJ/mol.

In order to check for potential effects brought about by the carrier gas, both $N_2$ and Ar were employed in separate experiments at 296 K. These measurements allowed us to examine the potential effects brought about by the presence of $C(^1D)$ atoms (generated by $CBr_4$ photolysis at the level of 10-15 %[29]) on the chemistry due to the relatively rapid quenching of $C(^1D)$ atoms by $N_2$.[33] Table 1 and Figure 4 clearly show that the derived rate constants at 296 K in Ar and $N_2$ are within the experimental error bars of the measurements, indicating the weak influence of the nature carrier gas. Indeed, as $C(^1D)$ atoms are also expected to react rapidly with $N_2O$[60] (rather than being quenched to the ground state) these atoms are not expected to interfere significantly with the kinetics of $C(^3P)$ loss. The room temperature rate constants measured here are seen to be slightly larger than those derived in the previous studies by Husain and Kirsch,[19] Husain and Young[20] and by Dorthe et al.[21], although the rate constant obtained by Husain and Kirsch[19] is essentially the same as the present one considering the experimental error bars of this earlier work.

Below 296 K, the measured rate constants for the $C + N_2O$ reaction increase as the temperature falls, reaching a value of $(7.87 \pm 0.80) \times 10^{-11}$ cm$^3$ s$^{-1}$ at 50 K; a factor of two larger than the room temperature value. A fit to the experimental data of the form $k(T) = \alpha(T/300)^\beta$ can be used to represent the temperature dependence of the rate constants. The fits yields values for $\alpha = (3.4 \pm 0.8) \times 10^{-11}$ cm$^3$ molecule$^{-1}$ s$^{-1}$ and $\beta = -0.50 \pm 0.05$ which are valid over the 50-296 K range. If we use these parameters to extrapolate the fit down to 10 K, a rate constant of $k_{C+N_2O}(10 \text{ K}) = 1.9 \times 10^{-10}$ cm$^3$ s$^{-1}$ is obtained.

The results of the present RRKM ME calculations are also shown in Figure 4, although there are no previous theoretical studies of the kinetics of the $C + N_2O$ reaction for comparison.

In common with the measurements, these rate constant values represent the overall loss of C($^3$P) + N$_2$O reagents, which could also potentially include stabilization of the C...N$_2$O complex rather than reaction to form products. Nevertheless, the ME calculations indicate that stabilization is entirely negligible in this shallow well (-6.7 kJ/mol with respect to the reagents corrected for ZPE) at all temperatures and at all pressures below atmospheric pressure. Indeed, the observed small variations in the calculated product yields originated from stochastic noise due to the finite number of simulations rather than real differences. Consequently, the derived rate constants are representative of the low-pressure limiting values. Moreover, as we obtain identical results for simulations performed with and without inclusion of tunneling effects, it seems unlikely that tunneling plays any role in the present reaction.

When a TS1 energy of -3.01 kJ/mol (corrected for ZPE) is employed in the ME simulations as derived by single point calculations at the RCCSD(T)-F12/AVTZ level, it can be seen from the green dotted line in Figure 4 that the calculated overall rate constants underestimate the measured values. At 300 K, we predict a rate constant $k_{C+N_2O}(300 \text{ K}) = 1.9 \times 10^{-11}$ cm$^3$ s$^{-1}$, almost a factor of two smaller than the measured values of $k_{C+N_2O}(296 \text{ K}) = (3.2\text{-}3.5) \times 10^{-11}$ cm$^3$ s$^{-1}$. In order to provide a better agreement between experiment and theory, we ran a series of simulations empirically adjusting the value of the TS1 barrier height. When the barrier is lowered to -5.85 kJ/mol with respect to the reagents, the calculated rate constants reproduce well the measured ones over the entire temperature range as shown by the red dashed line in Figure 4. This adjustment is well within the expected precision of the RCCSD(T) calculations which are usually considered to furnish energies with an accuracy approaching 1 kcal/mol (4.2 kJ/mol). Other factors such as vibrational anharmonicity could also affect the calculated rate constants, although only harmonic frequencies were considered here. Using this adjusted barrier height, at 10 K, we calculate a rate constant $k_{C+N_2O}(10\ K)$ = 1.7 × 10$^{-10}$ cm$^3$ s$^{-1}$, close to the calculated high-pressure limit rate constant at this temperature, $k_{capt}$, (the solid blue line in Figure 4) of 1.8 × 10$^{-10}$ cm$^3$ s$^{-1}$ obtained from a variational treatment of the barrierless entrance channel. Even without adjustment of the TS1 barrier height, the predicted rate constant at 10 K of 1.55 × 10$^{-10}$ cm$^3$ s$^{-1}$ is close to the $k_{capt}$ value.

**5.2 Product channels**

To the best of our knowledge, there are no direct previous theoretical studies of the product channels of the C + N$_2$O reaction. In related work, however, Zhu and Lin[61] performed an ab initio investigation of the NCN + O reaction which samples parts of the $^3A''$ potential surface of the NCNO intermediate involved in the C + N$_2$O reaction. They calculated that the spin-allowed exothermic product channels of this reaction over the triplet surface (N + NCO and CN + NO) do not present any barriers at the G2M (CC1) level. Moreover, their RRKM calculations (E, J-resolved) predicted that the products, CN + NO, should be preferentially formed, with the rate constant for this channel almost three orders of magnitude greater than the one for the N + NCO channel over the 200-3000 K temperature range. Although the reagents C + N$_2$O are approximately 100 kJ/mol higher in energy than the NCN + O ones, it is not expected that this will alter the preferred product channels, with CN + NO remaining as the dominant products.

In order to check for the possible formation of CO and N$_2$ as products we ran a series of additional experiments at 177 K. As the carrier gas used here was N$_2$, C($^1$D) atoms formed as the minor products of CBr$_4$ photolysis were rapidly quenched to the ground state and were not expected to interfere. Here, we detected the formation of atomic hydrogen by VUV LIF at 121.567 nm in a similar manner to earlier work,[31] following the introduction of a fixed concentration of ethylene, $C_2H_4$, into the reactor. $C(^3P)$ is known to react rapidly with $C_2H_4$ at room temperature and below[62] leading to the formation of H + $C_3H_3$ as the exclusive products.[63] Then, with $C_2H_4$ still present in the flow we added $N_2O$ to the flow. If the C + $N_2O$ reaction leads to the formation of CN + NO as exclusive products, then the CN radical products will also go on to react rapidly with $C_2H_4$ (present in excess with respect to CN), producing $C_2H_3CN$ + H as exclusive products.[64] Consequently, we would not expect to see a significant change in the H-atom signal intensity before and after the addition of $N_2O$ to the reactor. If CO and $N_2$ are produced in significant quantities by the C + $N_2O$ reaction then we would expect the H-atom signal intensity to be lower after addition of $N_2O$ to the reactor. The results of one of these experiments are shown in Figure 5.

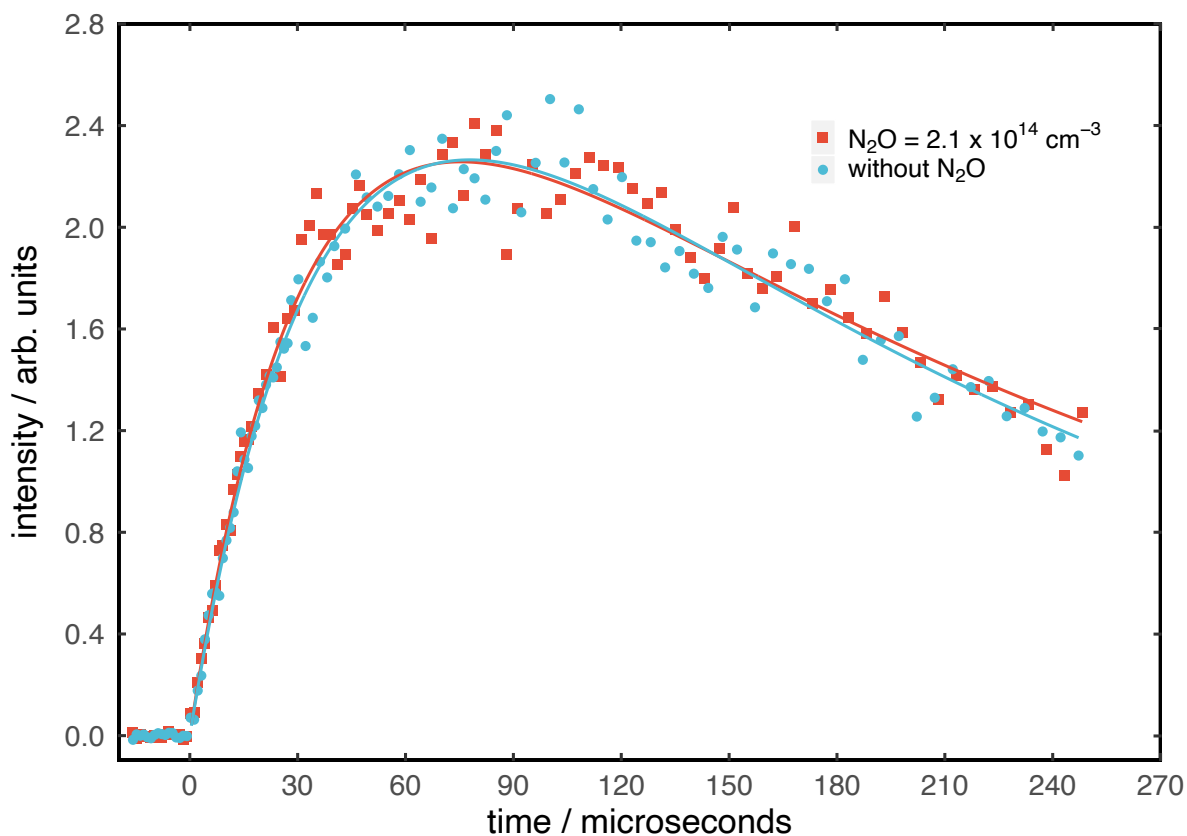

**Figure 5** H($^2$S) VUV LIF signal as a function of time recorded at 177 K in the presence of a fixed concentration of $C_2H_4$ (2.8 × 10$^{13}$ cm$^{-3}$). (Red solid squares) [$N_2O$] = 2.1 × 10$^{14}$ cm$^{-3}$ ; (light blue solid circles) without $N_2O$. Solid lines represent biexponential fits to the individual datasets of the form $A(\exp(-k_1't) - \exp(-k_2't))$.

A fixed concentration of $C_2H_4$ = 2.8 × 10$^{13}$ cm$^{-3}$ was used during these experiments (the light blue circles in Figure 5), leading to a pseudo-first-order rate constant for C-atom loss equal to approximately 9000 s$^{-1}$ ($k_{C+C_2H_4}(177\ K)$ = 3.2 × 10$^{-10}$ cm$^3$ s$^{-1}$). With $C_2H_4$ still present, $N_2O$ was added to the flow with a concentration of 2.1 × 10$^{14}$ cm$^{-3}$, leading to a pseudo-first-order rate constant for the loss of C-atoms equal to approximately 18000 s$^{-1}$ (here we considered $k_{C+N_2O}(177\ K)$ = 4.5 × 10$^{-11}$ cm$^3$ s$^{-1}$). These experiments are represented by the red squares in

Figure 5. If we compare the two formation curves, two different effects are apparent. (I) Although the loss rate for C-atoms is twice as large for the measurements recorded with N$_2$O present, which should lead to a faster H-atom formation rate, H-atom formation instead changes very little with and without N$_2$O. When both C$_2$H$_4$ and N$_2$O are present in the flow, the H-atom fluorescence signal produced directly by the C + C$_2$H$_4$ reaction is expected to be only half as large as the one produced by the C + C$_2$H$_4$ reaction alone, with the production of CN radicals expected to occur at the same rate. The remaining H-atoms formed by the CN + C$_2$H$_4$ reaction are then expected to be produced with a slower rate constant equal to approximately 9000 s$^{-1}$ ($k_{\mathrm{CN+C_2H_4}}$(177 K) = 3.2 × 10$^{-10}$ cm$^3$ s$^{-1}$) as the CN + N$_2$O reaction is negligible at low temperature. As a result of the delayed appearance of half of the atomic hydrogen signal, the overall effect is a H-atom appearance curve that is very similar to the one derived from the addition of C$_2$H$_4$ alone. (II) The H-atom signal intensity levels before and after the addition of N$_2$O are very similar. For the six pairs of curves recorded, the ratio of the peak values of the biexponential fit for those experiments with N$_2$O divided by those experiments without N$_2$O were all in the range 0.94 - 1.08, with a mean value of 1.01 ± 0.06 (errors quoted at the 95 % confidence level). Although we cannot establish quantitative product branching ratios based on the present analysis, this finding clearly indicates that the major products of the C+ N$_2$O reaction are CN + NO, while the CO + N$_2$ product channel is likely to represent at most a few percent of the total (and probably even lower).

**6 Astrophysical Model**

To test the effects of the C + $N_2O$ reaction on the chemistry of dense interstellar clouds, we introduced this reaction into the astrochemical gas-grain model Nautilus[65, 66] in its three-phase form.[67] Nautilus calculates the gas-phase and grain composition as a function of time, based on a modified version of the reaction network kida.uva.2014[68] assuming that the grain surface and mantle are both chemically active. The network comprises several thousand individual reactions both in the gas phase and on grains involving several hundreds of species, with the initial parameters as listed in Table 2.

**Table 2** Elemental abundances and other model parameters

| Species | Abundance[a] | $nH + 2nH_2$ / $cm^{-3}$ | T/ K | Cosmic ray ionization rate / $s^{-1}$ | Visual extinction |
|---|---|---|---|---|---|
| $H_2$ | 1.0 | $2 \times 10^4$ | 10 | $1.3 \times 10^{-17}$ | 10 |
| He | 0.18 | | | | |
| $C^+$ | $3.4 \times 10^{-4}$ | | | | |
| N | $1.2 \times 10^{-4}$ | | | | |
| O | $4.8 \times 10^{-4}$ | | | | |
| $S^+$ | $1.0 \times 10^{-5}$ | | | | |
| $Fe^+$ | $2.0 \times 10^{-8}$ | | | | |
| $Cl^+$ | $2.0 \times 10^{-9}$ | | | | |
| F | $1.3 \times 10^{-8}$ | | | | |

[a]Relative to $H_2$

Elements are either initially in their atomic or ionic forms in this model (elements with an ionization potential < 13.6 eV are considered to be fully ionized) and the C/O elemental ratio is equal to 0.7 in this work. More detailed descriptions of other parts of the model (such as those related to surface reactivity and desorption) can be found in Ruaud et al.[67]

Simulations were performed here with and without the inclusion of the C + $N_2O$ reaction, with a value calculated from the Arrhenius parameters derived above, to examine its influence on these species and the assumed products which are considered to be exclusively CN and NO based on the present analysis and electronic structure calculations as described above, as well as the previous work of Zhu and Lin.[61] Figure 6 shows the calculated relative abundances of CN, $N_2O$, NH and NO as a function of time obtained using the gas-grain model with the C + $N_2O$ reaction included (the nominal model) compared to those obtained in the absence of this reaction.

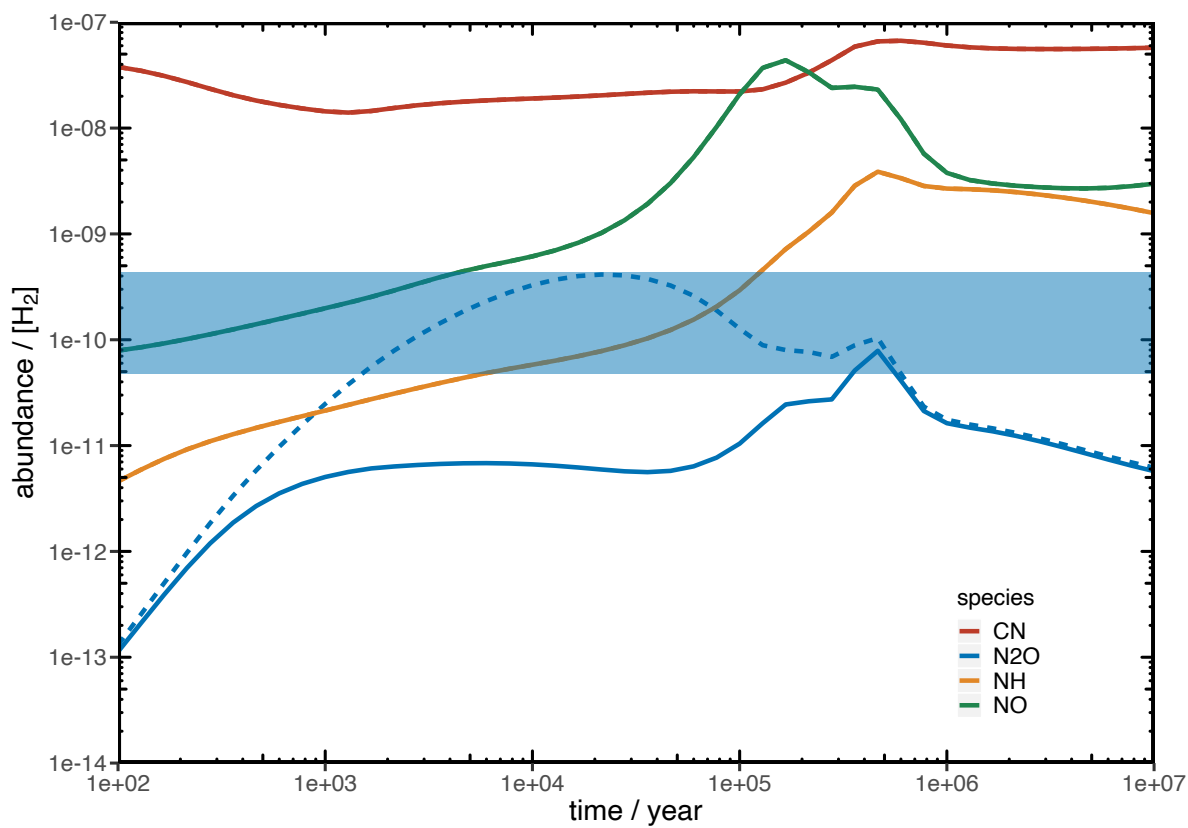

**Figure 6** Simulated gas-phase abundances of CN, $N_2O$, NH and NO relative to $H_2$ as a function of time. (Solid lines) nominal model (including the C + $N_2O$ reaction); (dashed lines) without the C + $N_2O$ reaction. Grain-surface reactions and gas-phase grain surface exchange processes are allowed for these simulations. The solid blue rectangle shows the observed $N_2O$ abundances from L483[6] with upper and lower limits corresponding to the observed abundance multiplied or divided by 3 respectively.

The simulations without the C + $N_2O$ reaction predict a peak value of $N_2O$ abundance reaching a few $\times$ $10^{-10}$ (with respect to $H_2$) around 2 $\times$ $10^4$ years. In current reaction networks there are very few efficient loss processes for $N_2O$. $N_2O$ is lost essentially through its interactions with cosmic rays and through ion-molecule reactions such as those with $He^+$, $C^+$, $H_3^+$ and $HCO^+$. Indeed, there are no efficient low temperature reactions of $N_2O$ with other neutral species included in current reaction networks. After inclusion of the C + $N_2O$ reaction in the network, the predicted $N_2O$ abundance falls by as much as two orders of magnitude between $10^3$ and $10^6$ years, as the large atomic carbon abundance at early times efficiently removes $N_2O$ from the gas-phase. In contrast, the abundances of product molecules CN and NO are mostly unchanged by the inclusion of this process, indicating that these species have other more important sources in the current model. After 6 $\times$ $10^5$ years, the simulated $N_2O$ abundance of both models is virtually identical. At this time, the gas-phase C-atom abundance is 100 times lower than its peak value (C is now mostly converted to CO) indicating that the C + $N_2O$ reaction is no longer the dominant removal mechanism for $N_2O$ (the dominant removal mechanism being the depletion of gas-phase $N_2O$ onto dust grains).

The comparison with observations is difficult because there is only one observation of $N_2O$ in a medium close to a dense cloud, the cold envelope around protostar L483, with a relative abundance of 1.45 $\times$ $10^{-10}$ with respect to $H_2$.[6] This value is approximately twice as large as the peak value for simulations with the C + $N_2O$ reaction included, which occurs around 5 $\times$ $10^5$ years. This result could indicate that reactions producing $N_2O$ are missing.

In order to identify the major sources for gas-phase $N_2O$ in the present model (gas-phase production, grain surface reactions or both), we also ran the model with the grain surface desorption mechanisms switched off (whilst still allowing species to deplete onto the grains), thereby calculating the $N_2O$ abundance through gas-phase formation routes alone. The results of these simulations are shown in Figure 7 and clearly demonstrate that grain production of interstellar $N_2O$ is important, with peak $N_2O$ abundances falling by almost an order of magnitude when desorption is prevented from occurring.

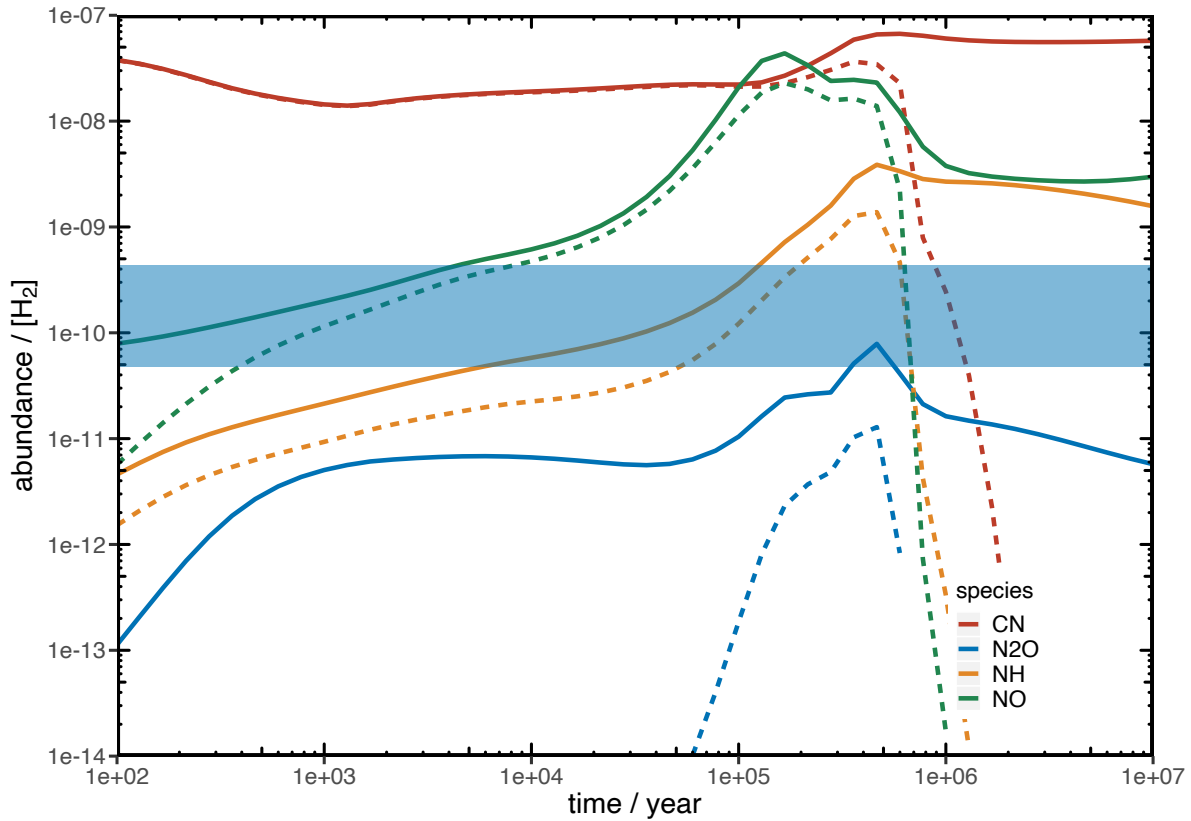

**Figure 7** Simulated abundances of CN, $N_2O$, NH and NO as a function of time. (Solid lines) nominal model (including the C + $N_2O$ reaction); (dashed lines) nominal model with desorption processes from the grain-surface switched off. The solid blue rectangle shows the observed $N_2O$

abundances from L483[6] with upper and lower limits corresponding to the observed abundance multiplied or divided by 3 respectively.

A closer look at the reaction network allows us to identify the NH + NO $\rightarrow$ H + N$_2$O reaction as the major gas-phase formation route for N$_2$O, with other N$_2$O sources involving the NO$_2$ molecule (the N, CN, NH$_2$ and NH + NO$_2$ reactions are all considered to form N$_2$O) being almost entirely negligible due to the extremely low predicted gas-phase NO$_2$ abundance. Even if NH and NO are mostly produced by gas-phase processes at times that correspond to typical ages of dense clouds ($10^5$ - few $10^5$ years), grain production sources of these species are far from negligible, for example by the chemical desorption of NH through the s-N + s-H $\rightarrow$ s-NH* $\rightarrow$ NH reactions (the prefix s- means species on grains). Consequently, a significant fraction of the N$_2$O produced in the gas-phase originates from NH and NO formed on the grains (and then released into the gas-phase).

Although the formation of N$_2$O by gas-phase reactions is not negligible, grain reactions are a much more important source in our model. The current simulations indicate that the majority of interstellar N$_2$O is formed on grains by two reactions: s-NH + s-NO and s-N + s-HNO. It should be noted that the s-N + s-NO reaction seems not to produce s-N$_2$O as shown by Minissale et al.[69] (the N + NO reaction has been studied theoretically in the gas-phase by Gamallo et al.[70] showing that it is a fast process giving products N$_2$ + O through an unstable intermediate: N$_2$O in a triplet state). In our network, the second reaction that efficiently (but indirectly) produces N$_2$O is the s-N + s-HNO $\rightarrow$ s-NH + s-NO reaction followed by recombination of s-NH + s-NO. There are no data for this reaction but preliminary calculations performed at the M06-2X/AVTZ level show that the barrier for abstraction of the hydrogen atom is low, if it exists at all, and should be efficient on interstellar ices due to tunneling as it involves a light hydrogen atom. The s-NH on grains and the s-NO formed by this abstraction are considered to recombine to form s-HNNO which will dissociate to give s-H and s-N$_2$O. Given the exothermicity of the NH + NO reaction and the high mobility of the hydrogen atom, s-H and s-N$_2$O are unlikely to recombine. These two reactions are efficient on interstellar grains because the binding energy of nitrogen atoms on water ice is assumed to be rather low (around 700 K)[71-73] giving it a certain mobility even at 10 K. It should be noted that our model reproduces quite well the abundance of N$_2$O in the protostar IRAS16293B,[1] where N$_2$O is produced on grains and then desorbs when the temperature of the protostar increases.

**7 Conclusions**

This paper presents the results of a joint experimental/theoretical/modeling study of the C($^3$P) + N$_2$O reaction at low temperature. Experimentally, a supersonic flow reactor was used to investigate this process over the 50-296 K range. Pulsed laser photolysis was used to generate C($^3$P) in-situ in the reactor, while these atoms were followed by pulsed laser induced fluorescence in the vacuum ultraviolet wavelength range. Theoretically, electronic structure calculations were performed on the ground $^3$A″ surface of CNNO, allowing the various intermediates, stationary points and possible product channels of the title reaction to be determined. Additionally, statistical rate theory was used to derive rate constants over the same temperature range, allowing the measured values to be extrapolated to lower temperatures and pressures. The reaction was seen to accelerate as the temperature falls, while supplementary measurements indicated that the major products are CN and NO with a negligibly small contribution from the CO and $N_2$ channel.

The new rate constants were introduced into a gas-grain model of dense interstellar clouds. These simulations predict that $N_2O$ abundances decrease significantly at intermediate times when gas-phase $C(^3P)$ abundances are high, while the predicted $N_2O$ abundances return to their previous levels at typical cloud ages when $C(^3P)$ is mostly locked up as CO or depleted onto interstellar grains.

**Author Information**

**Corresponding Author**

*Email: kevin.hickson@u-bordeaux.fr.

**Supporting Information**

Calculated geometries, energies, frequencies and rotational constants. Input parameters for the *ktools* calculations of $k_{capt}$.

**Acknowledgements**

K. M. H. acknowledges support from the French program ''Physique et Chimie du Milieu Interstellaire'' (PCMI) of the CNRS/INSU with the INC/INP co-funded by the CEA and CNES as well as funding from the ''Programme National de Planétologie'' (PNP) of the CNRS/INSU.

Supplementary information (SI) file for

**An Experimental and Theoretical Investigation of the Gas-Phase C($^3$P) + N$_2$O Reaction. Low Temperature Rate Constants and Astrochemical Implications**

Kevin M. Hickson,[1,*] Jean-Christophe Loison,[1] Pascal Larregaray,[1] Laurent Bonnet[1] and Valentine Wakelam[2]

[1] Univ. Bordeaux, CNRS, Bordeaux INP, ISM, UMR 5255, F-33400 Talence, France
[2] Univ. Bordeaux, CNRS, LAB, UMR 5804, F-33270 Floirac, France

C + N$_2$O theoretical calculations, geometries, rotational constants in GHz, frequencies and energies (distances in Angstrom, angles in degrees, frequencies in cm$^{-1}$ without corrections factors, energies in hartree). Parameters used in the *ktools* calculation of $k_{capt}$ (distances in Å, energies in kJ/mol, rotational constants in GHz, frequencies in cm$^{-1}$).

C:
**Energies (hartree):**
M06-2X/AVTZ:           -37.842846
CCSD(T)-F12/AVTZ:      -37.788374

| N$_2$O | 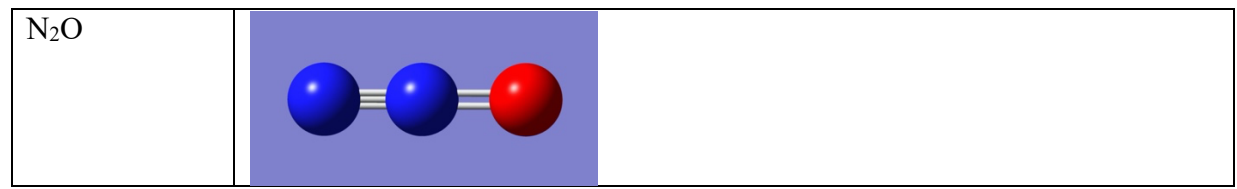 |
|---|---|

|       | M06-2X/AVTZ | CCSD(T)/aug-cc-pVQZ |
|-------|-------------|---------------------|
| N-N   | 1.1111      | 1.1290              |
| N-O   | 1.1796      | 1.1875              |
| N-N-O | 180         | 180                 |

**Rotational constant (GHZ):**
M06-2X/AVTZ:           12.8579
CCSD(T)-F12/AVTZ:      12.5772
Experimental[1]:       12.5616

**Frequencies (cm$^{-1}$):**
M06-2X/AVTZ:           653, 653, 1358, 2419
CCSD(T)-F12/AVTZ:      600, 600, 1301, 2284
Experimental[2]:       589, 589, 1285, 2224

**Energies without/with ZPE (hartree):**
M06-2X/AVTZ:           -184.6639352/-184.655022
CCSD(T)-F12/AVTZ:      -184.485008/-184.485008+0.010901

| | | |
|---|---|---|
| C…NNO | 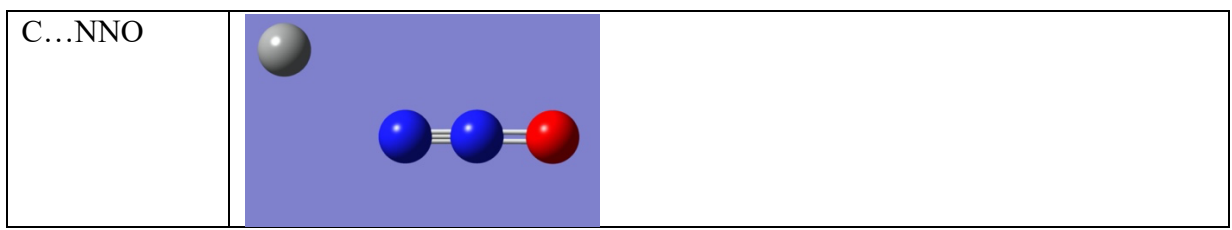 | |

| | M06-2X/AVTZ | CCSD(T)-F12/AVTZ |
|---|---|---|
| N-N | 1.1117 | 1.1287 |
| N-O | 1.1751 | 1.1848 |
| N-N-O | 179.48 | 180.02 |
| C-N$_2$O | 2.3079 | 2.5152 |
| C-N-NO | 144.37 | 157.07 |
| Diedre(C-N-N-O) | 180.00 | 180.00 |

**Rotational constant (GHZ):**
M06-2X/AVTZ:        3.5232    3.6398    109.9711
CCSD(T)-F12/AVTZ:   3.0476    3.0889    228.0196

**Frequencies (cm$^{-1}$):**
M06-2X/AVTZ:        22, 109, 646, 651, 1374, 2424
CCSD(T)-F12/AVTZ:   6, 92, 596, 597, 1311, 2297

**Energies without/with ZPE (hartree) (/C+N$_2$O (kJ/mol)):**
M06-2X/AVTZ:        -222.5137151/-222.501804 (-10.33)
CCSD(T)-F12/AVTZ:   -222.276175/-222.276175+0.011149 (-6.68)

| | | |
|---|---|---|
| TS1 | 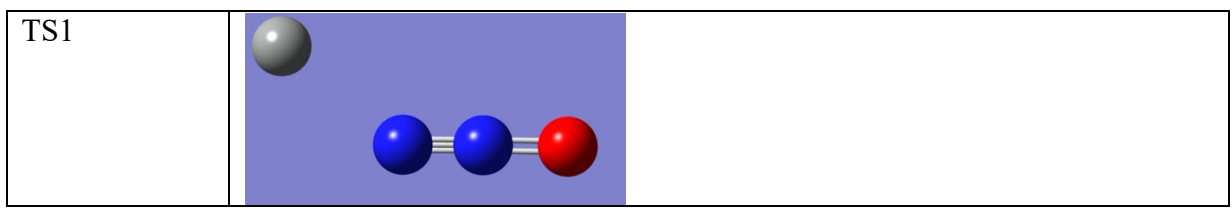 | |

| | M06-2X/AVTZ | CCSD(T)-F12/AVTZ |
|---|---|---|
| N-N | 1.1123 | 1.1306 |
| N-O | 1.1739 | 1.1810 |
| N-N-O | 179.39 | 178.70 |
| C-NNO | 2.1629 | 1.8955 |
| C-N-NO | 141.29 | 146.40 |
| Diedre(C-N-N-O) | 180.00 | 180.00 |

**Rotational constant (GHZ):**
M06-2X/AVTZ:        3.7629    3.9113.    98.4045
CCSD(T)-F12/AVTZ:   4.0941    4.2067    152.8735

**Frequencies (cm$^{-1}$):**
M06-2X/AVTZ: 121i, 8, 641, 650, 1375, 2421
CCSD(T)-F12/AVTZ: 233i, 76, 558, 574, 1301, 2298

**Energies without/with ZPE (hartree)/C+N$_2$O (kJ/mol):**
M06-2X/AVTZ: -222.5136467/-222.502039 (-10.95)
CCSD(T)-F12/AVTZ: -222.27457640/-222.27457640+0.01094918(-3.01)

| TS2 | 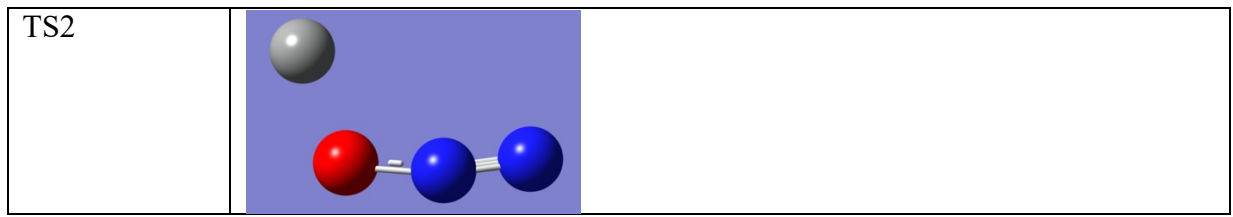 |
|---|---|

|  | M06-2X/AVTZ |  |
|---|---|---|
| N-N | 1.1142 |  |
| N-O | 1.2638 |  |
| N-N-O | 160.40 |  |
| C-ONN | 1.5274 |  |
| C-O-NN | 114.07 |  |
| Diedre(C-O-N-N) | -62.93 |  |

**Frequencies (cm$^{-1}$):**
M06-2X/AVTZ: 1180i, 193, 555, 583, 887, 2197

**Energies without/with ZPE (hartree)/C+N$_2$O (kJ/mol):**
M06-2X/AVTZ: -222.4993404/-222.489282 (+22.54)

| TS3 | 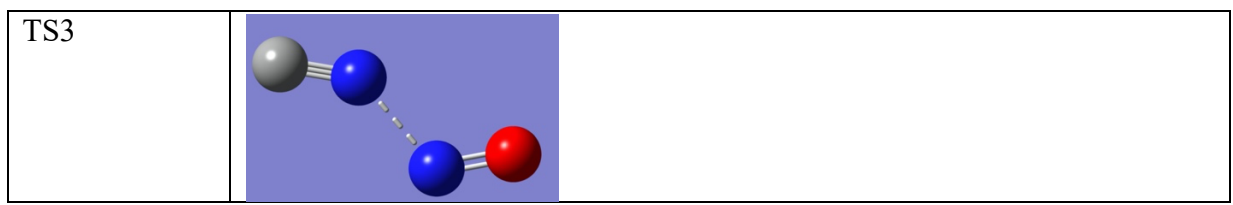 |
|---|---|

|  | M06-2X/AVTZ |  |
|---|---|---|
| N-N | 1.7801 |  |
| N-O | 1.1548 |  |
| N-N-O | 119.88 |  |
| C-NNO | 1.1813 |  |
| C-N-NO | 140.11 |  |
| Diedre(C-N-N-O) | 180.00 |  |

**Frequencies (cm$^{-1}$):**
M06-2X/AVTZ: 655i, 103, 155, 339, 1813, 1989

**Energies without/with ZPE (hartree)/C+N$_2$O (kJ/mol):**
M06-2X/AVTZ:     -222.5736285/-222.563609 (-172.60)

| TS4 | 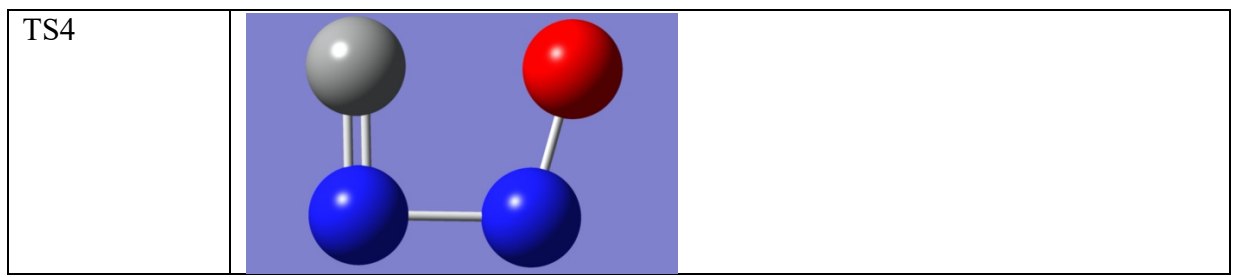 |
|---|---|

|  | M06-2X/AVTZ |  |
|---|---|---|
| N-N | 1.4380 |  |
| N-O | 1.2798 |  |
| N-N-O | 104.79 |  |
| C-NNO | 1.2385 |  |
| C-N-NO | 91.95 |  |
| Diedre(C-N-N-O) | 0.00 |  |

**Frequencies (cm$^{-1}$):**
M06-2X/AVTZ:     601i, 462, 614, 735, 1256, 1656

**Energies without/with ZPE (hartree)/C+N$_2$O (kJ/mol):**
M06-2X/AVTZ:     -222.5492746/-222.538518 (-106.73)

**Table S1** Parameters employed in the *ktools* calculation of $k_{capt}$ derived at the CCSD(T)-F12/AVTZ level.

| Distance / Å | energy ZPE corrected kJ/mol wrt reagents | energy ZPE uncorrected kJ/mol wrt reagents | A / GHz | B / GHz | C / GHz | ν1 / cm$^{-1}$ | ν2 / cm$^{-1}$ | ν3 / cm$^{-1}$ | ν4 / cm$^{-1}$ | ν5 / cm$^{-1}$ | ν6 / cm$^{-1}$ |
|---|---|---|---|---|---|---|---|---|---|---|---|
| 7 | -0.146765553 | -0.169869969 | 50.88714 | 0.80047 | 0.78807 |  | 4 | 600 | 600 | 1301 | 2284 |
| 6 | -0.244959322 | -0.292218355 | 67.28242 | 1.01417 | 0.99911 |  | 7 | 600 | 600 | 1301 | 2285 |
| 5 | -0.550305186 | -0.621456286 | 73.22642 | 1.3432 | 1.31901 |  | 11 | 600 | 600 | 1301 | 2285 |
| 4 | -1.658266962 | -1.777202196 | 67.70668 | 1.8808 | 1.82997 |  | 17 | 600 | 600 | 1302 | 2286 |
| 3.5 | -3.100455123 | -3.126185041 | 101.5403662 | 2.1682417 | 2.1229102 |  | 20 | 599 | 600 | 1303 | 2286 |
| 3.4 | -3.437832109 | -3.474589135 | 72.8943646 | 2.3151533 | 2.2438865 |  | 24 | 599 | 600 | 1304 | 2287 |
| 3.2 | -4.336278839 | -4.375661367 | 76.3563317 | 2.4965466 | 2.4175039 |  | 26 | 599 | 600 | 1305 | 2287 |
| 3 | -5.409321441 | -5.609647231 | 114.5815414 | 2.6168761 | 2.5584449 |  | 30 | 599 | 600 | 1306 | 2290 |
| 2.9 | -6.059920797 | -6.132384648 | 261.8346715 | 2.6199768 | 2.5940205 |  | 37 | 599 | 600 | 1307 | 2292 |
| 2.8 | -6.129759146 | -6.616002087 | 270.0641349 | 2.725465 | 2.6982346 |  | 66 | 599 | 600 | 1308 | 2293 |
| 2.7 | -6.505731009 | -7.014290716 | 279.1178349 | 2.836807 | 2.8082653 |  | 71 | 598 | 598 | 1309 | 2294 |
| 2.6 | -6.608913232 | -7.17707183 | 91.3057894 | 3.1666439 | 3.0605004 |  | 80 | 597 | 600 | 1310 | 2292 |
| 2.5152 | -6.682689833 | -7.331713888 | 228.0195791 | 3.0888778 | 3.0475934 | 6 | 92 | 596 | 597 | 1311 | 2297 |